# Casimir interaction of excited media in electromagnetic fields

Yury Sherkunov

## Introduction

The long-range electric dipole interaction between an excited atom and a ground-state atom is considered in ref. [1,2] with the help of perturbation theory.

The result for the interaction potential between two dissimilar atoms is as follows

$$U(R) = -\frac{4}{9\pi} \int_0^\infty \frac{\omega_A \omega_B |d_{eg}^A|^2 |d_{eg}^B|^2}{(\omega_A^2 + u^2)(\omega_B^2 + u^2)} u^6 e^{-2uR} \left[ \frac{1}{u^2 R^2} + \frac{2}{u^3 R^3} + \frac{5}{u^4 R^4} + \frac{6}{u^5 R^5} + \frac{3}{u^6 R^6} \right] du \quad (1.1)$$
$$- \frac{4}{9} \frac{|d_{eg}^A|^2 |d_{eg}^B|^2 \omega_A^6}{(\omega_B^2 - \omega_A^2)} \left[ \frac{1}{\omega_A^2 R^2} + \frac{1}{\omega_A^4 R^4} + \frac{3}{\omega_A^6 R^6} \right] \theta(\omega_A).$$

Here $\omega_A$ and $\omega_B$ are transition frequencies of the excited atom A and the ground-state atom B correspondingly, $d_{eg}$ is the matrix element of dipole moment, R is the distance between the atoms, $\theta(\omega_A)$ is the unite step-function, which signifies that the second term of the right-hand side of the equation is not equal to zero only for the case of excited atom A. If the atoms are in their ground state the second term equals zero. If the distance R between the atoms is less than the wavelength of atomic transitions R<<λ – the van der Waals case – the resonant term of Eq. (1.1) reads

$$U(R) = -\frac{4}{3} \frac{|d_{eg}^A|^2 |d_{eg}^B|^2 \omega_A^6}{(\omega_B^2 - \omega_A^2) R^6}. \quad (1.2)$$

For the Casimir-polder case of large distances (R>>λ), the Eq.(1.1) is as follows

$$U(R) = -\frac{4}{9} \frac{|d_{eg}^A|^2 |d_{eg}^B|^2 \omega_A^4}{(\omega_B^2 - \omega_A^2) R^2}. \quad (1.3)$$

The formulas (1.2) and (1.3) correspond to resonance interaction between the atoms. The interaction can be either attractive or repulsive depending on frequency detuning of the atoms.

If both atoms are in their ground state, the equations interaction potentials differs from (1.2) and (1.3). For the van der Waals case we will have the London formula [3]

$$U(R) = -\frac{2}{3} \frac{|d_{eg}^A|^2 |d_{eg}^B|^2}{(\omega_B + \omega_A) R^6}. \quad (1.4)$$

For the Casimir-Polder case we will have [4]

$$U(R) = -\frac{23}{4\pi R^7} \alpha_A(0) \alpha_B(0), \quad (1.5)$$

where $\alpha_A(0)$ and $\alpha_B(0)$ are the polarizabilities of the atoms at zero frequencies. We should stress here that the difference between the Eqs. (1.5) and (1.3) is significant. The potential of the interaction between two ground-state atoms drops as $1/R^7$ with the distance, while the potential of the interaction between excited and ground-state atoms drops as $1/R^2$.

In this paper we are going to investigate the interaction between two dissimilar atoms one of which is excited. We will consider the interaction between an excited atom and medium of dilute gas. It will be shown that the perturbation theory is not sufficient in this case. Direct implementation of the perturbation technique results in divergence of integrals. We will show that the interaction is suppressed due to absorption of photons by the medium. To take into account the absorption we use a non-perturbative method offered in Ref. [5]. Then we will consider the interaction between two media of dilute gases under thermal equilibrium. The results of calculations will be compared with the Lifshitz formula. We will demonstrate the vilotion of the Lifshitz formula for the temperature high enough to possess excited atoms.

## Interaction between an excited atom and a dielectric surface
## Perturbation method

Here we will consider the Casimir-Polder interaction (R>>λ) between an excited atom A and a gas media using perturbation technique. If the medium is diluted enough we can take into account pair interactions only. In this case the potential can be found by integrating the equation(1.3) (Fig.1).

$$U = -\int_V \frac{4}{9R^2} \frac{|d_{eg}^A|^2 |d_{eg}^B|^2 \omega_A^4}{(\omega_B^2 - \omega_A^2)} n dV = -\frac{8\pi n |d_{eg}^A|^2 |d_{eg}^B|^2 \omega_A^4}{9(\omega_B^2 - \omega_A^2)} \int_{z_0}^{\infty} dz \int_0^{\infty} \frac{\rho d\rho}{(\rho^2 + z^2)} = \infty \quad (1.6)$$

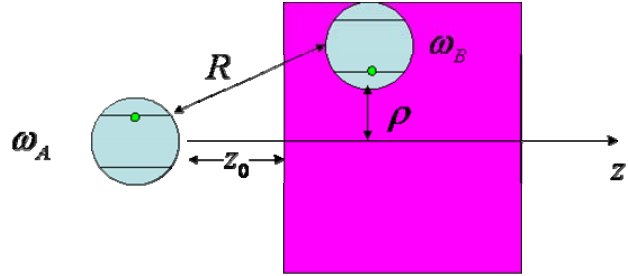

**Fig.1**

The interaction potential (1.6) is divergent. It means that one can not use perturbation method to find the interaction potential for an excited atom near a dielectric surface for the Casimir-Polder case.

## Interaction between an excited atom and a dielectric surface
## Non-perturbative approach

Here we will use the method of quantum Green's functions (Ref.[5]) to take into account the absorption of photons in the medium.

The Hamiltonian of the system is as follows

$$\hat{H} = \hat{H}_A + \hat{H}_B + \hat{H}_{med} + \hat{H}_{ph} + \hat{H}_{int},$$

$$\hat{H}_{int} = -\int \hat{\psi}^\dagger(r-R_A) \hat{d}^\nu \hat{E}^\nu(r) \hat{\psi}(r-R_A) dr - \int \hat{\varphi}^\dagger(r-R_B) \hat{d}^\nu \hat{E}^\nu(r) \hat{\varphi}(r-R_B) dr$$

$$-\int \hat{\chi}^\dagger(r-R_m) \hat{d}^\nu \hat{E}^\nu(r) \hat{\chi}(r-R_m) dr$$

$$\hat{\psi} = \sum_i \psi_i(r-R_A) \hat{b}_i, \hat{\varphi} = \sum_i \varphi_i(r-R_B) \hat{\beta}_i,$$

$$\hat{H}_{ph} = \sum_{k\lambda} \omega(\lambda) \left( \hat{\alpha}_{k\lambda}^{\dagger} \hat{\alpha}_{k\lambda} + \frac{1}{2} \right)$$

$$\hat{H}_A = \sum_i \varepsilon_{Ai} \hat{b}_i^{\dagger} \hat{b}_i,$$

$$\hat{H}_B = \sum_i \varepsilon_{Bi} \hat{\beta}_i^{\dagger} \hat{\beta}_i$$

Firs we will consider a system of two atoms embedded in a dielectric medium. The interaction potential can be found as the energy shift of one atom, say atom B, due to the presence of the other atom

$$U(\mathbf{R}) = \Delta E_B = Re\left[ M_{11}^{00}(\varepsilon_{B0}) \right]$$

Where $M_{11}^{00}(\varepsilon_{B0})$ is the mass operator of atom B [5].

The density matrix is

$$\rho_{12}^B(x,x') = \langle \hat{T}_c \hat{\psi}_1(x) \hat{\psi}_2^{\dagger}(x') \hat{S}_c \rangle, \text{ with scattering matrix } \hat{S}_c = \hat{T}_c \exp\left\{ \sum_{l=1,2} (-1)^l i \int_c \hat{H}_{intl}(t) dt \right\}.$$

After calculations and standard substitution $\frac{1}{2} \to \left( N_{k\lambda} + \frac{1}{2} \right)$, which takes into account the thermal photons, we find

$$U = U_{nr} + U_r,$$

$$U_{nr} = -2T \sum_{n=0}^{\prime} \alpha_g^{v_1 v_2}(i\zeta_n) \alpha_e^{vv'}(i\zeta_n) \zeta_n^6 e^{-2\zeta_n R} \left( \frac{1}{\zeta_n^2 R^2} + \frac{2}{\zeta_n^3 R^3} + \frac{5}{\zeta_n^4 R^4} + \frac{6}{\zeta_n^5 R^5} + \frac{3}{\zeta_n^6 R^6} \right), (1.7)$$

$$\zeta_n = 2\pi n T$$

$$U_r(R) = -\frac{2}{9} Coth\left( \frac{\omega_A}{2T} \right) \frac{|d_{eg}^A|^2 |d_{eg}^B|^2 \omega_A^6 (\omega_B - \omega_A)}{(\omega_B - \omega_A)^2 + \left(\frac{\gamma_B}{2}\right)^2}$$

$$\times \left[ \frac{1}{\omega_A^2 R^2} + \frac{1}{\omega_A^4 R^4} + \frac{3}{\omega_A^6 R^6} \right] \exp\left( -\frac{\gamma_{ph} R}{2} \right) \theta(\omega_A) \quad (1.8)$$

The life-time of the photon for the case of dilute gas medium is $\tau_{ph} = \gamma_{ph}^{-1} = \frac{3\gamma}{8\pi\omega |d|^2 n}$.

The non-resonance term (1.7) coincides with the one obtained in paper [6] for the interaction between two ground-state atoms. The resonance term generalizes the Power formula (1.3) for the case of two atoms in an absorbing medium. Now one can see that if the formula (1.8) is substituted into (1.6) the result will be divergent no more.

### Interaction between two media of excited atoms

Here we will consider the interaction between two media of dilute gases separated by a distance L>>λ.

To find the force we will integrate the equations (1.7) and (1.8) over the volume. To simplify the calculations we will use the following model. We will omit the exponent in (1.8), but the width of the interacting media will be restricted by the free pass of a photon in these media

$$L_{ph} = c\tau_{ph} = c\gamma_{ph}^{-1} = \frac{3c\gamma}{8\pi\omega|d|^2 n}.$$

After integration we should differentiate the potential with respect to L to find the force.

$$F(L) = F_{Lif}(L) + \frac{8\pi\hbar|d_A|^2|d_B|^2 \omega_A\omega_B(\omega_B^2 - \omega_A^2)n_An_B}{9c^4\left((\omega_B^2 - \omega_A^2)^2 + (\gamma_B\omega_A)^2\right)\left(1 + exp\left(-\frac{\omega_A}{T}\right)\right)\left(1 + exp\left(-\frac{\omega_B}{T}\right)\right)}$$

$$\left(\omega_A^3 e^{-\frac{\omega_A}{T}} coth\left(\frac{\omega_A}{2T}\right) - \omega_B^3 e^{-\frac{\omega_B}{T}} coth\left(\frac{\omega_B}{2T}\right)\right)\left[(2L_{ph} + L) Log\left(\frac{2L_{ph} + L}{L_{ph} + L}\right) - L \cdot Log\left(\frac{L_{ph} + L}{L}\right)\right].$$

(1.9)

The Lifshitz formula for the same case of high temperatures reads [7]

$$F_{Lif}(L) = \frac{2\pi T|d_A|^2|d_B|^2}{9L^3 \omega_A\omega_B} tan\left(\frac{\omega_A}{2T}\right) tan\left(\frac{\omega_B}{2T}\right) n_A n_B. \quad (1.10)$$

In Fig.2 we represent the dependence of the normalized force on the distance L. We see that the dependence obtained with the help of quantum electrodynamics results in either attraction or repulsion depending on the distance between the media.

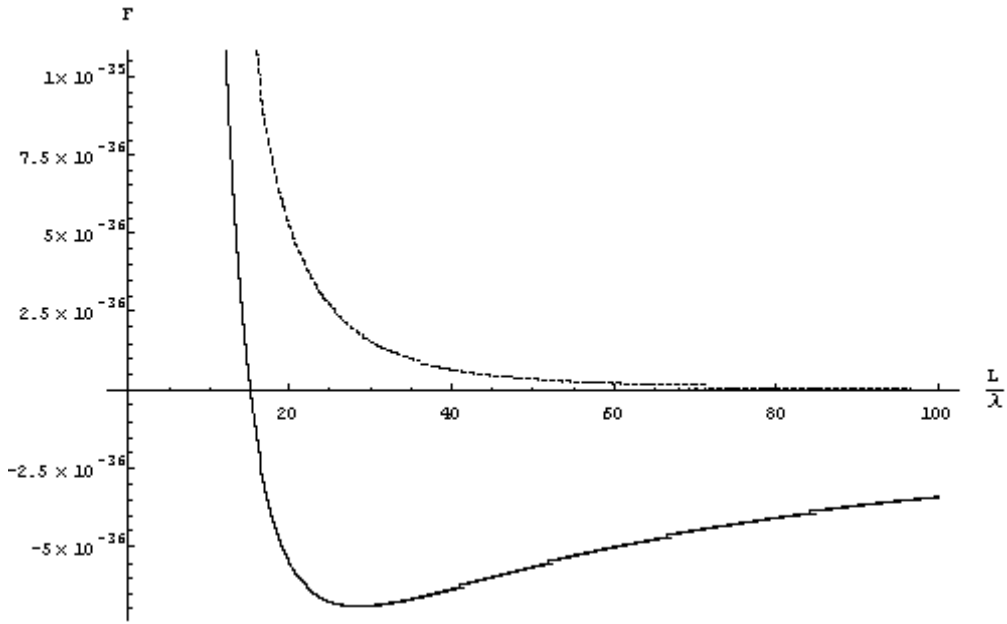

Fig.2 The dependence of the normalized force on distance. The upper curve corresponds to the Lifshitz formula (1.10), the lower curve corresponds to Eq. (1.9). ($\omega_A = 1.1\omega_B, T = 0.4\hbar\omega_B$)

The temperature dependences of the forces are given in Fig.3 and Fig.4.

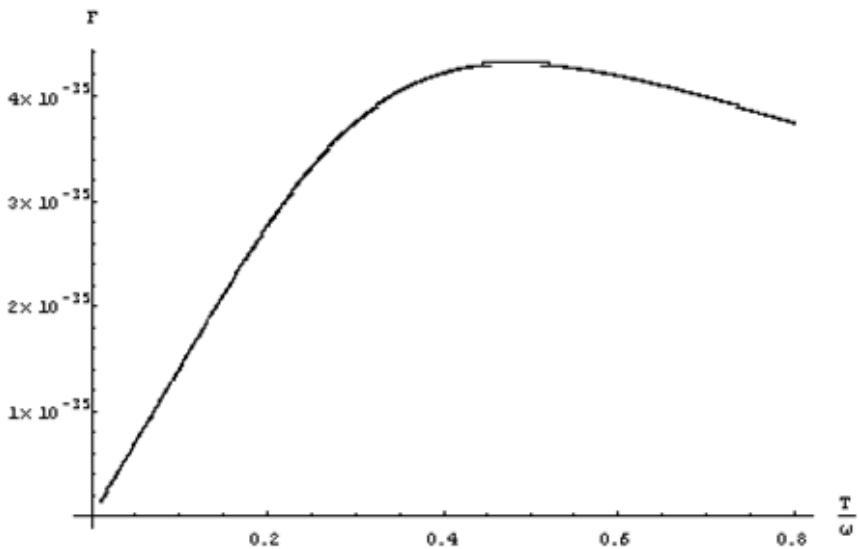

Fig.3 The typical temperature dependence of the force obtained with the help of the Lifshitz formula (1.10)

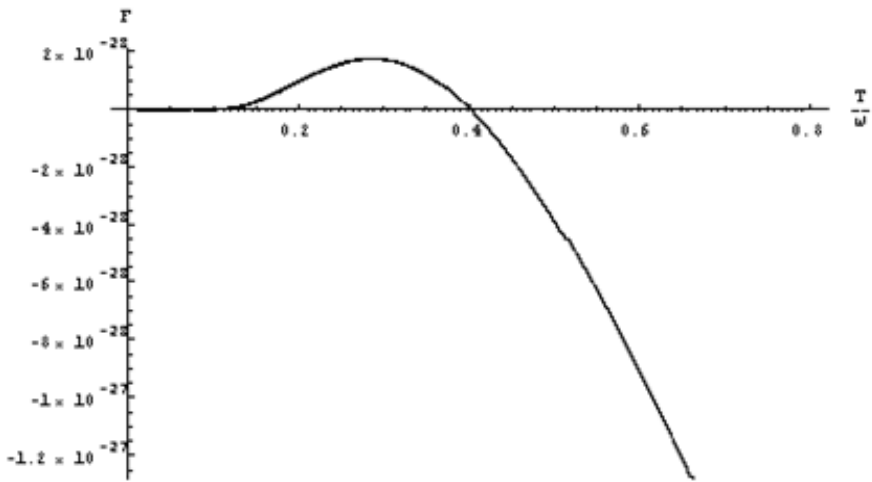

Fig.4 The typical temperature dependence of the force obtained with the help (1.9)

## Conclusions

We considered Casimir-Polder interaction between an excited and a ground-state atom. We showed that the perturbation theory results in a divergence in the case of interaction between an excited atom and a gas medium. Using non-perturbative method, we found the interaction potential between two atoms in a dielectric medium. The suppression of interaction due to absorption of photons in the medium is demonstrated. It results in convergence of the integrals for an excited atom interacting with a medium. Then we considered a case of Casimir interaction between two gases. The results of quantum electrodynamics are compared with the Lifshitz formula.